# Towards Launching AI Algorithms for Cellular Pathology into Clinical & Pharmaceutical Orbits

Amina Asif[†], Kashif Rajpoot[‡], David Snead[§], Fayyaz Minhas[†], Nasir Rajpoot[†§¶*]

[†]Tissue Image Analytics Centre, Department of Computer Science, University of Warwick, UK
[‡]Department of Computer Science, University of Birmingham, UK
[§]Department of Pathology, University Hospitals Coventry & Warwickshire, UK
[¶]The Alan Turing Institute, UK
[*]Corresponding Author: Nasir Rajpoot. Tel: +44(24)7657-3795; Email: N.M.Rajpoot@warwick.ac.uk

## Summary

Computational Pathology (CPath) is an emerging field concerned with the study of tissue pathology via computational algorithms for the processing and analysis of digitized high-resolution images of tissue slides. Recent deep learning based developments in CPath have successfully leveraged sheer volume of raw pixel data in histology images for predicting target parameters in the domains of diagnostics, prognostics, treatment sensitivity and patient stratification – heralding the promise of a new data-driven AI era for both histopathology and oncology. With data serving as the fuel and AI as the engine, CPath algorithms are poised to be ready for takeoff and eventual launch into clinical and pharmaceutical orbits. In this paper, we discuss CPath limitations and associated challenges to enable the readers distinguish hope from hype and provide directions for future research to overcome some of the major challenges faced by this budding field to enable its launch into the two orbits.

## 1   Background

The field of Computational Pathology (CPath) has been growing in interest among academic, industrial and healthcare sectors over the recent years. The surge in CPath literature, industrial outfits (startups and big tech/pharma alike) and potential healthcare applications has been spurred by the development of advanced machine learning (ML), and particularly deep learning (DL), based algorithms. These methods consume tissue images with associated clinical metadata and/or annotations and promise to offer assistance to practicing histopathologists, oncologists and clinicians in accurate and efficient diagnosis and treatment planning for cancer patients. A major advantage of such algorithms is their being objective and reproducible; contrast this with a pathologist's opinion based on visual examination that is inherently subjective and may lead to intra- and inter-observer variability. The coming together of the digitization of tissue slides, advances in DL and accessibility of high-performance computing hardware have resulted in a rapidly growing number of DL algorithms for a broad variety of pathology tasks (1) showing state-of-the-art (SOTA) performance, sometimes claiming to surpass *human* performance. Examples include cell detection and classification, mitosis detection, cancer detection and segmentation, cancer classification and subtyping, genetic expression prediction and mutation status prediction (2–11). High accuracy figures reported in the literature for these and other related tasks can be seen as evidence of the immense potential of DL algorithms for successfully modeling CPath problems. Large investments in the form of public and private funding add to the promise of achieving CPath breakthroughs with DL technologies, with consequential and likely impact on the practices of pathology and oncology.

Several excellent review articles including (12–17) have recently covered CPath research trends from different angles. We only present a brief overview of recent research trends[1] in CPath. Publications recorded on PubMed and WoS, respectively show a 20-fold and 15-fold increase in ML/DL based CPath research activity for the period 2010-2020. The most frequently studied tasks are classification and segmentation of tissues and cells. In recent years, there has been increased interest in prediction of gene expression and mutation status with the availability of large datasets like Cancer Genome Atlas (TCGA), cBioPortal and METABRIC. In terms of cancer sites, breast cancer has received the most attention, followed by prostate and colorectal cancers, perhaps owing to the prevalence of these cancer types while stomach, thyroid and pancreatic cancer have received the least attention. Supervised learning is the most commonly employed learning paradigm; however, due to annotation-related limitations (discussed in more detail later), other paradigms including weakly supervised, semi-supervised, self-supervised and unsupervised learning have been gaining interest among CPath researchers. Supplementary Figure 1 presents a graphical summary of this analysis. To summarize, a broad spectrum of ML techniques, with DL dominating the field currently, have been applied to model a number of prediction tasks over the recent past. However, the rise in popularity of DL in general and DL for CPath in particular comes with an associated hype that can lead to somewhat unreasonable expectations from the models and inadvertent and potentially serious consequences in a clinical setting due to adoption of substandard technologies without appropriate scrutiny.

Despite the publication of several peer reviewed studies reporting extraordinary performances, a key challenge in the clinical uptake of CPath technologies is that they may not generalize well to new/unseen datasets and therefore, may be not be ready for launch into the clinical orbit. In addition, several other challenges in different phases of development including, but not limited to, scarcity of publicly available datasets and models, lack of stringent and problem-specific performance evaluation protocols, absence of uniform standards and regulatory policies and lack of reproducibility of methods can be regarded as being responsible for impediment in development of effective models for CPath. Several characteristics of CPath data and problems such as the giga-scale of images, tissue and tumor heterogeneity within and across patients and patient subgroups, scarcity of accurate annotations and susceptibility to overfitting due to high-dimensional data necessitate the development of robust solutions that can help overcome fragility in models introduced by these problems. We present an in-depth analysis of limitations and challenges associated with four main phases of the development of a CPath system with an aim to identify factors impeding effective integration of CPath systems in clinical workflows. We conclude this paper with recommendations for addressing the challenges and list open problems for future research in the field.

## 2   Limitations, Challenges and Recommendations

NASA's Technology Preparedness Levels (18) is already established as a means of assessing the readiness of new technologies for prime-time use. Extending this theme, we identify limitations and challenges associated with four main phases of the AI development lifecycle in CPath using the analogy of rocket launching: data (*fuel*) collection and curation, model (*engine*) development, robust performance evaluation (*rigorous testing*) and deployment (*orbital launch*), as shown in **Figure 1**. We conjecture that overcoming some of these limitations and addressing the challenges will be essential to making CPath systems suitable for effective integration in both research and clinical workflows. We also make recommendations on how to address some of these challenges.

---

[1] To conduct this research trend analysis, we searched for computational pathology and digital pathology published papers which have image analysis, artificial intelligence (AI) or machine learning (ML) relevant contribution using Clarivate Analytics' Web of Science (WoS) and PubMed. The search was conducted via Boolean operators with search query component matches found in title, abstract, or keywords to keep our search focused to relevant papers. The literature search was conducted on 6th November 2021.

## 2.1 Data (*The Fuel*)

Collection and curation of datasets is the first and most important step in any ML study. In the absence of good quality datasets that are true representatives of the respective populations, it is very hard to develop effective models and perform realistic performance assessment. In a CPath study, the dataset may typically contain whole slide images (WSIs) of tissue slides together with the associated clinical and genomic annotations. The most commonly used approach is to deal with the imaging data in the same way as in other computer vision (CV) tasks that typically deal with natural images. However, CPath images (ie, WSIs) differ from data in other domains including CV in several ways, posing the following data-related challenges. It is striking that the most popular sources of this "data fuel" (The Cancer Genome Atlas or TCGA) is derived from projects established for very different objectives, namely genomics. Hence it is not surprising this data fuel is of insufficient "grade" to develop effective models and perform realistic performance assessment. We discuss some of the challenges specific to CPath in terms of data collection, curation and processing as follows.

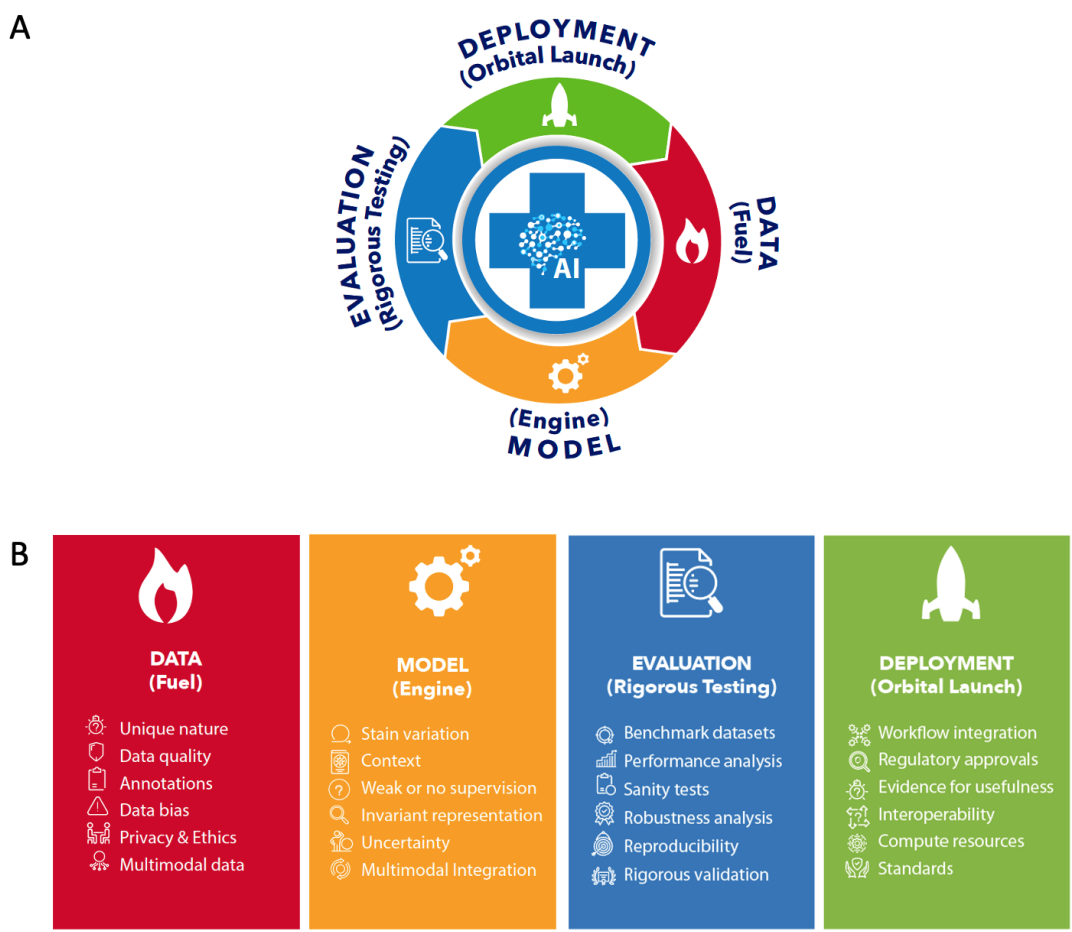

*Figure 1: CPath challenges and limitations associated with the four main phases of the AI development lifecycle, using the analogy of rocket launch: Data (Fuel), Model (Engine), Evaluation (Rigorous Testing) and Deployment (Launch).*

### 2.1.1 Uniqueness of CPath datasets

CPath data typically consists of whole-slide images (WSIs) of tissue sections with associated clinical and genomic annotations. A WSI is typically multi-gigapixel in size, hence leading to large storage and processing requirements (19). However, despite individual WSIs being giga-scale, WSI datasets usually contain only a relatively small number of independent examples – i.e., these datasets are typically tall and thin (20). Scarcity in term of number of independent examples combined with the problem of high dimensionality in CPath makes DL models highly prone to overfitting (21).

DL models have been shown to be *data hungry*, i.e., they require large datasets for training to generalize well on unseen data (22). This problem can be mitigated to some extent by learning low-dimensional representations but the need for sufficiently large datasets stands nonetheless (24,25). One of the biggest factors contributing to the success of deep models in generic CV applications for natural images is the availability of large-scale datasets like ImageNet, CIFAR, COCO (23–25). Natural images are relatively easier to collect, store and label since these images do not typically require specialized equipment or highly trained personnel for data acquisition and annotation. In contrast, WSIs require specialized scanning equipment and trained technical staff for preparation and scanning of the slides. Furthermore, annotations and target labels at the cell, regional or slide levels by trained pathologists may be needed for training purposes. Unlike natural images, a typical WSI is typically multi-gigapixel in size, hence leading to larger storage and processing requirements (19). TCGA has played a big role in development of CPath algorithms by offering publicly available WSIs along with genetic and clinical annotations (26). More such initiatives with large-scale publicly available datasets are required for developing effective DL solutions.

Apart from differences in acquisition, dimensionality and scale, CV and CPath datasets contrast in the characteristics of image contents as well. Tissue landscapes differ a lot from natural images. While objects-of-interest in CV datasets are usually easily identifiable with well-defined boundaries and not a lot of intra-image heterogeneity, objects in tissue slide images, i.e., cells and different tissue regions are relatively difficult to discern, may be large in number (eg, there may be tens to hundreds of thousands of cells in a tissue slide) and suffer from both inter- and intra- class heterogeneity (27). To develop effective solutions, all these characteristics should be taken into account while developing ML models for CPath, especially if transfer learning is employed (28).

### 2.1.2 Data quality

The time-tested adage '*garbage in, garbage out*' applies to WSIs too. Although there have been significant advances in digital slide scanning technology, image quality challenges arising from issues with the nature and quality of tissue sections mounted on glass slides (such as tissue folds, slide vendor labels, pen markings, mounting issues, stain quality and age related issues) and those arising from the technology itself (such as missing faint tissue and out-of-focus tissue contents) remain. As a pre-processing step, quality checks (QC) need to be routinely performed on the WSIs to detect and mitigate image quality issues.

### 2.1.3 Data annotation and discordance

Several supervised learning tasks require precise cell-level or region-level annotations for training and evaluation purposes (29). These annotations are expensive, tedious and time-consuming. Furthermore, unlike natural images, where techniques like crowdsourcing can be used to acquire labels, trained personnel may be needed to annotate histopathology images accurately. In a recent study, Amgad et al. (30) demonstrated that crowdsourcing from medical students for annotating cell nuclei in breast cancers was considerably accurate, hence providing a relatively more efficient framework for data annotation. However, the effectiveness of crowdsourcing for other types of annotations and more complex problems still remains to be evaluated.

Another related issue is the discordance among pathologist labels due to the inherent subjectivity of visual assessment (31–33). Discordant annotations, in addition to being a source of labeling noise, lead to a disagreement over what to be used as ground truth for supervised training and performance evaluation. Consensus results of labeling by multiple pathologists (34) could be used as more reliable ground-truths for such tasks, often incurring additional cost to the process of annotation.

### 2.1.4 Data biases

Biased data leads to biased ML/DL models which can have serious implications. Like other data dependent fields, CPath models can be adversely affected by biases in data. Several types of biases can be encountered in CPath datasets and studies introducing different challenges. We discuss some of these as follows.

#### 2.1.4.1 *Cancer types/subtypes*

Most of the recent CPath research is focused on applying DL to specific tissue sites. It has been established that because of their ability to model complex decision surfaces, DL requires large amounts of data to avoid overfitting and generalize well for a task (35,36). Due to such large data requirements, cancer types that are relatively rare tend to be neglected in CPath research (37). As mentioned above, even though there are more than 100 different types of cancers (38), CPath research studies are mostly focused on only a handful of cancer types. This not only hinders research leading to development of computational models for rarer cancers, but also impedes new discoveries related to such cancer types.

Initiatives involving collection and curation of datasets for rarer cancer types are needed for developing better models and better evaluation of computational methods. Bigger datasets for rarer cancers are also needed for effective pan-cancer studies as well.

#### 2.1.4.2 *Ethnicity*

Like other healthcare informatics data, CPath data is highly prone to ethnic bias (39–42), referred to as the *health data poverty* problem (43). Despite significant technological advancements, developing and under-developed parts of the world still lack sufficient infrastructure for generating digital histopathological and genomic data leading to lack of representation from these regions. Population under-representation combined with the fact that histology and prevalence of cancers can vary highly across different races (44–46) not only raises concerns about universal generalization of models but also increases the gap in access and applicability of advanced tools and solutions to poorly represented ethnic groups. To overcome this issue, initiatives for data collection from underrepresented ethnicities are needed. Furthermore, CPath models need to be tested and corrected for potential racial biases (47).

#### 2.1.4.3 *Statistical biases*

Survival analysis and predictive modeling of treatment effects using histopathology and -omics data are two of the most widely studied problems in CPath. Cancer survival datasets, like other survival related datasets, are vulnerable to immortal time bias (ITB) and selection bias (48). ITB is the manifestation of using inaccurate start times in studying effects of a variable. For example, using date of enrollment in a study as the starting point may not be appropriate in analyzing the effect of a drug if there is a significant time gap in its administration and enrolment date. It has been shown that such biases can lead to significant over/under-estimation of treatment efficacy and effects of other covariates of interest (49).

The immortal time and selection biases need to be identified and corrected for effective modeling. In CPath survival studies, it is not a common practice to evaluate and correct for ITB, sometimes because only the times of enrollment in studies is available and accurate information about treatment start times is not available. For studies like treatment effect prediction, datasets with sufficient number of examples and accurate times to events with corrected immortal time bias are needed.

### 2.1.5 Privacy and ethical challenges

Privacy concerns over healthcare data may be regarded as the biggest cause of scarcity of publicly available datasets. A related challenge is the implication of a participant's right to unenroll from a study at any time. Access revoking may require a patient's data to be removed not only from *all* related databases but also from *all* models that are trained using their data, i.e., making the models *unlearn* an example. There has been some work in developing unlearning techniques for models but effective and time-saving solutions are still needed and this is still an open problem in ML (50,51). AI in healthcare brings with it a number of ethical concerns; for instance, the potential to aggravate discrimination and inequality due to biased ML algorithms (52). Comprehensive coverages of ethical challenges in incorporating ML models for healthcare and ethical issues in pathology are presented in (53) and (54), respectively. Furthermore, the approach to sharing of medical data for the greater public good may need to be revisited, especially in the European context where "AI based on European values" may be too slow-moving (55).

### 2.1.6 Multimodal data collection and curation

Recently, there has been a surge of interest in developing prognostic and predictive DL models for patients using WSIs (14). While there may be some important signals in a WSI, expecting it to provide the complete representation of a patient's expected survival is not realistic, as histopathology data is just a small part of a big complex picture. In clinical practice, diagnosis and treatment regimens for cancer patients are not based on histopathology reports alone. Doctors use a patient's history, underlying health conditions, reports from other tests and functional imaging from other modalities in addition to histopathology data for suggesting treatments and future course of action. That is, histopathology data is just a small part of a big complex picture. For effective modeling of such problems, additional information like genomic and clinical data are needed, which indicates the necessity for systems built on multimodal data. However, collecting, curating and collating multimodal datasets is far from straightforward. In addition to challenges associated with collaboration, correspondence and data sharing among different centers for creating such datasets, another issue is that not all types of data may be available for all patients, leading to a significant number of missing entries in a dataset and necessitating the development of specialized modeling techniques with support for heterogeneous and missing data for downstream analysis.

## 2.2 Modeling (*The Engine*)

The goal of ML models in CPath is to learn a suitable representation of tissue morphology and architecture associated with detection and recognition of disease group/phenotype, molecular genotype, treatment effects, other omics signatures and important objects in a tissue slide (for example, objects such as cell nuclei, microvessels and tubules may also need to be delineated). In this section, we discuss some challenges specific to modeling in CPath.

### 2.2.1 Modeling stain characteristics

Most existing approaches fail to model the domain-specific characteristics of images in CPath and treat them as *natural RGB* images. Such approaches do not explicitly model the fact that WSIs are obtained through a multi-step process that has a significant impact on their characteristics. Variations in tissue processing steps such as chemical fixation or freezing, dehydration, embedding and staining can change the visual characteristics of the tissue slide in a non-uniform manner across tissue types and laboratories well before the tissue slide is scanned to produce WSIs. One important aspect of this process that is not modelled explicitly in existing computational pathology approaches is staining. Pathology images are obtained by staining a given sample with one or more dyes that absorb incident light depending upon dye concentrations and their binding characteristics for different histological components (e.g., proteins, DNA, etc.). Routine stains such as Hematoxylin, Eosin and special stains for antibody-optimized immunohistochemical (IHC) markers are often used for pathology diagnosis and/or biomarker analysis. Consequently, the image acquisition process in routine histopathology is predominantly based on light absorption in these stains. This contrasts with natural images obtained through standard digital cameras, such as those in the widely used ImageNet database, which are unstained and operate on a different lighting and camera model involving reflected, absorbed, and radiated light. Typically, "stain variations" in the tissue are corrected post hoc by stain estimation, normalization or augmentation approaches to generate RGB images which are then used as input to DL pipelines for solving various ML problems. Failure to explicitly model stain absorption characteristics and their associated non-linearities across tissue can have a significant impact on the generalization performance (56).

### 2.2.2 Context and multi-resolution nature of WSIs

Another major challenge in computational pathology is explicit modelling of the multi-resolution nature of whole slide images. Pathologists typically analyze histological patterns at different magnification levels for visual assessment inevitably taking into account the contextual information for their decision-making. This highlights the importance of visual context when analyzing low-level details of the histological landscape and combine diagnostic evidence in a multiresolution manner. Due to the sheer size of WSIs, a WSI is often divided into image tiles (or *patches*) at a specific magnification, making the problem of modeling context in WSIs more challenging than it may be for natural images. Training and inference are both typically performed with limited context captured by individual patches, with the underlying assumption being that each patch is an *independent* data point. In addition, CPath algorithms also face the well-known *signal-frequency uncertainty* dilemma: the broader the context, the less precise the localization of a region or object. A multi-resolution approach can integrate predictive information at multiple levels, at the cost of increase in model complexity potentially also requiring more training data for effective learning. Another compromise is a distributed attention mechanism that can integrate information across multiple spatial locations and magnification levels. Existing methods in computational pathology have attempted to address these challenges to some degree (57). However, to the best of our knowledge, no existing method has demonstrated its ability to model context effectively across a variety of computational pathology tasks.

### 2.2.3 The case for weak or no supervision

The size of WSIs poses a major problem in the form of computational bottlenecks in performing gradient computations while training DL models. Most existing CPath methods employ patch-level analysis, assuming that the patch labels are available and can provide a direct supervisory signal for effective training. However, obtaining patch-level labels can be challenging and typically only WSI-level labels are available for training, making the case for so-called *weak supervision*. Weakly supervised CPath algorithms such as (6,35,58–61) aggregate patch-level prediction scores by different mechanisms such as majority voting, average pooling or multiple instance learning. The success of these approaches depends on the nature of the ML task and validity of assumptions underlying these approaches. Recently, self-supervised learning methods such as (62,63) that exploit supervisory signals in the data itself with the help of domain-specific as well as domain-agnostic tasks have been shown to be successful for effective tumor detection with limited annotations. However, development of a truly generalizable weakly supervised or self-supervised approaches remains an open problem.

### 2.2.4 Learning invariant representations

ML methods in CPath require an effective representation of input images that are robust to extraneous variations resulting from factors like rotation, translation, slide preparation and staining, scanner characteristics etc, in order to allow the model to generalize well to unseen test data. The invariances can be learned through various augmentation strategies, self-supervised learning (62,63) and contrastive learning (64). In addition to the symmetries associated with classical images such as translation and rotation, CPath models also need to cater for domain-specific invariances, including invariances associated with *technical* changes such as stain and scanner characteristics as well as histological properties underlying a prediction task. For example, variations in breast tissue density or fat content across population types can impact tumor subtype classification models. Such variations, if not factored in the development of CPath models, can lead to generalization failure. Although several approaches have modeled technical invariances, explicitly modelling histological variations in CPath models and learning domain-specific invariant representations remain largely unexplored.

### 2.2.5 Modeling uncertainty and out of domain (OOD) detection

Modeling label uncertainties in model training and generating uncertainty (or confidence) scores with inference are key requirements for practical utility of CPath models. Confidence scores can enable predictive models to "*know what they don't know*", detect OOD test examples and abstain from generating a decision in such cases (65,66). A few existing approaches have addressed this issue (67,68). However, this dimension of CPath model development requires further attention for their use in practice.

### 2.2.6 Integration of data from multiple modalities

Development of integrated solutions that utilize data from multiple sources such as radiology, pathology images, genetic sequencing and transcriptomics, multi-spectral and multiplexed imaging, spatial transcriptomics, clinical data, clinical letters and lab reports is an open area of research in computational diagnostics. Mining such data can reveal interesting associations and lead to the discovery of novel biomarkers and early diagnosis of multiple diseases. Some approaches have been proposed for the fusion of patho-radiomic and patho-genomic features (69,70). However, in order to model such solutions as ML problems, a key challenge is the availability of linked multimodal datasets. As a consequence, approaches such as learning using privileged information that assume that data from some modalities may only be available during training but not during inference can be very helpful. Development of such models requires a close interaction between national and potentially international health providers and ML researchers. One solution may be to provide an anonymized public data exchange that can accelerate the development of such solutions.

## 2.3 Performance Evaluation (*Rigorous Testing*)

DL models in CPath with their promise of enhanced efficiency and accuracy herald the dawn of data-driven AI era for the practice of cellular pathology in clinical and pharmaceutical workflows. Their deployment in practice, however, requires stringent performance evaluation as the decisions produced by these models are expected to have implications on patients' health and drug discovery roadmaps. In conventional settings, ML researchers attempt to estimate the ML model's accuracy on unseen data by using cross-validation protocols and testing on independent sets (71). However, models may still not generalize well to unseen data (72), often due to lack of robust performance evaluation. Below we cover some of the limitations and challenges concerning realistic performance evaluation and rigorous validation of CPath models.

### 2.3.1 Lack of available benchmark datasets

One of the biggest hurdles in accurate performance assessment of CPath models is the shortage of openly available, high quality and broadly representative benchmark datasets (73), leaving researchers with no choice but to evaluate their models over data that might be conveniently available but may not be a good representation of the real-world. For fair evaluation and comparison of methods, benchmark datasets should capture characteristics of real test data '*from the wild*' with a sufficient number of examples following the expected test data distribution and ideally representing all segments of the population. A benchmark dataset should also follow the FAIR principle of data management (74), i.e., it should be findable, accessible, interoperable and reusable.

### 2.3.2 Experiment design

While conducting performance evaluation of a model, the most important part is to ensure that experiment design is appropriate for realistic and reliable performance evaluation. Several factors can cause over-/under-estimation of expected performance of a method (75). Below, we discuss some ideas for experiment design that can lead to a fair and reliable performance assessment of a model.

### 2.3.2.1 Baseline analysis and fair comparison

In CPath, a number of methods based on complex, multi-stage strategies using DL methods have been proposed over the past several years. A significant number of such studies lack comparison with simpler methods. To solve a problem, baselines comprising simple methods should be established before applying complex techniques. This provides evidence that the observed performance improvement cannot be achieved with simpler techniques and hence, justifies the added complexity to an approach.

Furthermore, while performing comparison among different methods for solving a problem, it should be ensured that the experimental conditions are kept same for all the methods. This includes using same data examples, same level of hyperparameter optimization and not fixing splits that favor one method over the other. Factors that may favor one method over the other should be identified and resolved. Comparison among different methods can be regarded fair only if same level of parameter optimization is performed for all of them.

### 2.3.2.2 Indirect use of test sets

In traditional settings, a ML method is developed with hyperparameters tuned using a validation set and then tested on a test set. If the method does not perform well enough, it is then improved and tuned further so as to produce better results over the test set. Tuning parameters and hyperparameters until an improvement is observed in the test set may introduce overfitting to the dataset even though it is not being used directly (76). Improvement observed in such cases, i.e., multiple training cycles until the performance over the test sets improves, can also be a manifestation of false discovery phenomenon due to multiple testing. That is, a method trained in such a manner may show good performance on a particular test set but that may not be due to good generalization of the method and hence may fail to produce acceptable results in unseen and real-world data.

To prevent this phenomenon of achieving good test accuracy as a fluke due to multiple testing, the test set for a problem should be used only once, and reuse of test sets should be discouraged. If the method does not perform well and retuning of parameters is performed, additional unseen data should be used in testing. This, however, can be challenging due to the scarcity of data caused by factors explained in detail in section 2.1.

### 2.3.2.3 Evaluation metrics and interpretability

Choosing the right evaluation metric for gauging the performance of a system is important. Biases and imbalances in data, and their expected effect on the results should be considered before selecting evaluation metrics for a study, e.g., class imbalance may lead to over/under-estimated performance in terms of accuracy and AUC-ROC in classification problems. Furthermore, it should be checked if an evaluation metric is suitable for a particular problem, i.e., to ensure it quantifies performance in terms of the trait of interest in a system. When employing statistical tests, the underlying assumptions regarding distribution of data and the applicability of tests should be checked.

In addition to analyzing a method in terms of accuracy-based performance metrics, analyzing what features a model learns is important to avoid spurious correlations being used as discriminative features. One of the biggest hurdles in model interpretability in DL is their black-box nature. Several methods have been proposed in the literature to visualize feature maps in a model (77,78).Furthermore, techniques like SHAP, LIME can be used to analyze a model for feature importance (79,80). Feature map analysis should be performed to get an idea of what type of features are being learnt during training. This can act like first line of defense against spurious correlations as discriminative features.

While analyzing performance and features in a model, it is important to consider their significance and meaningfulness for domain experts. That is, how the numbers translate to metrics understandable and useful for the intended users of a system. For instance, in a system developed for cancer screening, an expert might be more interested in analyzing the occurrence of false negatives instead of just overall accuracy. The premise here being that the false negatives are worse errors in this case as compared to false positives as someone being falsely diagnosed with cancer will be identified later in the diagnostic pipeline. However, false negatives will lead to silent errors which have a lesser chance of being corrected and are expected to have dire consequences for patients.

#### 2.3.2.4 Train/Test splitting and stratification

It is a common practice to divide a dataset into training and testing subsets for the purpose of performance evaluation. While performing train/validation/test splitting in a dataset, class ratios should be maintained the same across the subsets, i.e., data should be stratified according to classes. This would help avoid over-/under-estimation of performance by maintaining the same data distribution in train and test samples.

Another factor that can cause over-estimation of performance results in CPath models is the patient-level overlap in train and test samples. Extending the argument further, *broad validation* consisting of unseen test data from external centers should be preferred to *narrow validation* where unseen data from the same center can be used for testing purposes (81).

### 2.3.3 OOD and sanity tests

Digitized WSIs of tissue slides often require cleaning up and removing of irrelevant and noisy regions like pen markings, background and other artifacts. In practice, CPath models can encounter WSIs with artifacts as well as out-of-distribution (OOD) WSIs. The model should be able to tell OOD samples apart from noisy images and images of interest. There is scarcity of research on developing models that can *abstain* from prediction for examples that are either too noisy or do not belong to the distribution of interest.

### 2.3.4 Robustness analysis

Typically, cross-validation and independent set validation are used for evaluating performance of a model. For further analysis of what a model is learning, gradient based visualization techniques like saliency maps are employed. Given the saliency maps light up in area of interest and validation produces high accuracy scores, a model is regarded as a suitable choice for solving a problem.

DL models have been shown to be vulnerable to adversarial attacks and small perturbations (82). Even highly accurate models may lack robustness to small variations and fail miserably. Though adversarial attacks are less likely for healthcare models, small perturbations are quite probable due to variations in factors like staining, scanning environments and equipment (83,84). Therefore, cross-validation and independent set testing, though necessary, may not be sufficient for performance assessment. *Fragility analysis* to evaluate how a model would respond to changes is required in CPath and other healthcare applications. A model should be deemed deployable only if demonstrates adequate robustness to small changes in inputs.

One of the ways to quantify strength of a model is by analyzing its response to addition of random noise, well-curated attacks, and variations in color. If there is a large change in its output despite a small imperceptible change in the input, the model should be considered fragile, and steps should be taken to improve its robustness. Techniques like adversarial training, data augmentation have been shown to build some immunity to adversarial attacks and improve generalization (85,86).

### 2.3.5 Reproducibility and repeatability

Major reproducibility crisis is being faced in several scientific domains (87,88) including ML in general and its application to healthcare in particular. There is a large number of methods with SOTA accuracies being reported frequently in the literature with a significant fraction that cannot be reproduced or repeated because of several factors. Two major causes of the lack of reproducibility in CPath are: unavailability of data and methods, often citing privacy concerns or due to commercial conflicts, and missing or incomplete preprocessing details. For

CPath in particular, this includes information regarding data preparation, quality check measures for WSIs, discarded examples/cases, color normalization techniques, patch extraction and selection etc. To ensure successful reproducibility, details like model initialization techniques, data augmentation, batch sizes, hyperparameters, data splits are needed. Missing these details can lead to problems in successful replication of results. To handle reproducibility and repeatability issues in CPath, recommendations in (89) inspired from the FAIR principles (74) can be followed.

## 2.4 Deployment (*Orbital Launch*)

The ultimate aim of a CPath algorithm is often to automate and assist with the pathologist's assessment of tissue slides. Additionally, CPath methods can also be employed for deep mining and discovery of novel histological patterns for prognostic and predictive biomarkers. Either way, in order to ensure that CPath algorithms are launched and stay in the clinical and pharmaceutical orbits, CPath systems must consider the following aspects of deployment.

### 2.4.1 Workflow integration

CPath solutions should ideally be integrated into the existing clinical and pharmaceutical workflows in order to automate or assist the pathological decision-making processes. Careful integration with existing laboratory information management (LIM), electronic health record (EHR), image management (IM) systems and/or trial databases may appear to be low-tech problems but is crucial for seamless workflow in routine pathology and oncology practice. Launching a separate CPath application, a common paradigm followed by several current CPath solution providers, that runs side-by-side all the above systems can only be the second-best option.

### 2.4.2 Validation and regulatory approvals

A CPath solution that can be deployed in routine clinical practice needs to have been validated rigorously to generate clinical evidence required for confidence of and buy-in from clinicians in the solution. Most healthcare systems require the solution to have also passed regulatory approvals, such as the Food and Drug Administration (FDA) in the United States and In Vitro Diagnostics (IVD) in the European Union. Going forward, CPath algorithms may need stringent regulatory approvals as they become more autonomous (15), for instance the issue of where the responsibility will lie in case an autonomous CPath algorithm goes completely awry. This need is further highlighted by aforementioned challenges associated with reliability and robustness.

### 2.4.3 Evidence for usefulness

Before a practical CPath solution can be deployed in practice, there should be sufficient robust evidence for its usefulness in terms of efficiency gains, higher accuracy, cost savings etc. Typically, well-designed health economic studies are required to generate evidence for efficiency gains and cost savings. Lack of such evidence may hamper the wider buy-in from pathology and oncology community and may also make it difficult for the laboratory or hospital management to justify investment in deployment of the solution, given the relatively high initial setup cost of the digital and computational pathology infrastructure.

### 2.4.4 Interoperability and lack of standards

There is some evidence to suggest that DL algorithms do not perform equally well on images from different scanners or even different versions of the same scanners. CPath solutions must demonstrate interoperability for various types of WSI formats generated by different slide scanners in order to help ensure that they are able to deal with this particular source of variation that is known to result in *domain shift* and are not biased towards or against pixel data from one or more image formats. Standardization of output formats for decisions made and annotations done by CPath algorithms will further enable interoperability of algorithms and aid with workflow integration. It is hoped that international industry-academic-clinical cooperative efforts for finalization of interoperability standards (such as the WSI DICOM standard) will help address these challenges.

### 2.4.5 Hardware requirements

Two major infrastructure related challenges facing the deployment of CPath models in routine workflows are: (a) the WSIs are large in their size and hence require large storage and high transfer bandwidth and (b) CPath models are often compute intensive. In this context, at least the following three models have emerged off late: (a) the **cloud** based data-to-compute model whereby WSIs are typically shipped to and processed in the cloud; this model offers the attractive feature of *pay-per-use* options without requiring significant compute-heavy investment but may give rise to potential data sharing and privacy concerns, (b) the **central** compute-to-data model, whereby data is shipped into a central repository and various compute solutions are brought over to be executed within the repository environment; this model is attractive for central repositories and for users without access to compute and storage resources but is likely to incur high initial setup cost, and (c) the ***federated*** learning model, whereby the DL model is trained locally without having to share the data, a global model is put together by merging the local models and then the global model is shared with all the contributing sites; a slight caveat of this model is that it requires sufficiently powerful compute resources at all contributing sites to be able to train local models. Going forward, further research on studying the pros of cons of these models in the context of CPath and further technological advances in computing, storage and networking may lead to appropriate balance between hardware costs and data sharing concerns.

### 2.4.6 Deployment timeline and the spectrum of mundanity

An oft-asked question is: which AI apps will be available for practical use in the next *x* years? To answer this question, we would like to refer the reader to **Figure 2** that we term as the *spectrum of mundanity*. On the one end of the spectrum, we have somewhat mundane tasks like driving a car to take the example of daily life or in the case of a pathology problem, finding tumor in a biopsy or lymph node or counting the number of mitotic cells – the kind of objective problems that the pathologists can easily solve with high accuracy and reproducibility. On the other end of the spectrum, there are quite obscure tasks pathologists are unable to deliver to anything more than a *gut feeling*; for instance, predicting a specific mutation from visual examination of a histology slide. And in the middle, there are relatively hard tasks (often reflecting complex interplay between tumour and host which may escape human observation at least to any reproducible manner) that require large amounts of data for algorithm development and large-scale prospective multi-centric validation with long follow-up; examples are risk scoring for malignant transformation, local recurrence or distant metastasis of cancer and prediction of response to a particular therapy. We conjecture that CPath solutions for tasks on the two ends of the spectrum that match (left) or surpass (right) the pathologist performance are the ones that will be deployed in routine practice sooner than those in the middle.

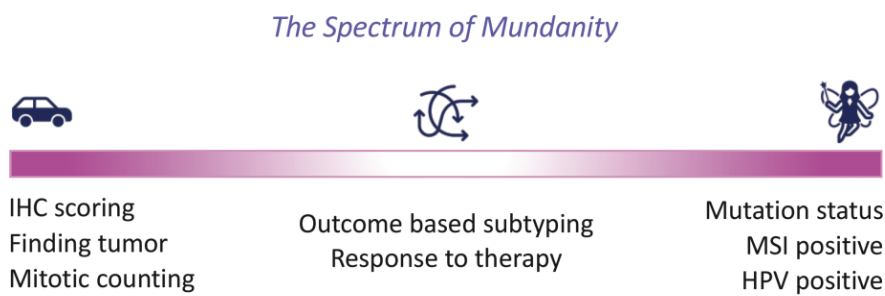

*Figure 2: Complexity of CPath tasks in terms of their mundanity.*

## 3   Conclusions and Future Directions

The nascent field of CPath offers significant promise to learn known histological patterns and mine unknown cellular and tissue architectural motifs with potentially practice-changes applications to cellular pathology based diagnostics, prognostics, treatment selection and patient stratification. There have been various positive developments in DL based CPath in recent years, showing great promise for facilitating enhancement in pathological assessment of tissue slides. However, some challenges remain to be addressed to make the vast majority of CPath methods truly generalizable and applicable in practice.

In the previous section, we have made recommendations regarding some of the challenges associated with the four phases of the AI development lifecycle. To conclude this paper, we discuss some research directions and open questions in the following sub-sections.

### 3.1 Data Modalities

Recent developments in imaging and transcriptomics offer great promise as future research avenues in CPath. Research in combining genetic information with histopathology, for instance using spatial transcriptomics, to study the impact of gene activity over cellular morphology and tissue architecture can lead to new discoveries and better understanding of cancer development, progression and treatment effects (90). Computational modeling using newer modalities including multiplexed images and MUSE imaging (91,92) and volumetric imaging for 3D modeling of tissues (93) are other emerging areas in CPath. However, as mentioned earlier, sufficiently sized multi-centric repositories need to be set up for effective modeling especially when using DL.

### 3.2 Frozen Tissue Images

Most computational pathology approaches focus on developing models for images obtained by scanning Formalin-Fixed Paraffin-Embedded (FFPE) tissue sections. However, cryotomy is routinely required for rapid diagnosis of surgical specimens as well as for enzyme histochemistry and spatial transcriptomics applications. Development of effective ML models using frozen tissue sections is challenging due to generally poorer image quality which complicates detection of tissue and cellular morphological patterns. Despite the clinical significance of these approaches, development of ML models for frozen tissue images remains relatively unexplored.

### 3.3 WSI holistic analysis

The predominant DL strategy for WSI analysis employs distributing the WSI into patches, perform patch-based model development, and aggregate the results from patches to complete WSI level analysis. This comes at the cost of losing neighborhood and multi-resolution context which is very important in a pathologist's analysis. In addition, many studies randomly choose patches to manage computational burden. While feeding the WSI to an algorithm may be a distant goal, there is a need for the development of analytical approaches that incorporate patch neighborhood and multi-resolution context during model learning and estimate high-level patch saliency to select salient patches for patch-based analysis from WSI.

### 3.4 Extraction of subvisual insights

The multi-gigapixel WSI data undoubtedly contains rich source of information about tissue micro-architecture. On the other hand, the pathologist's analysis is constrained by time and human visual system capabilities and thus may miss out on extracting the subvisual insights present in the tissue micro-architecture. For future developments in computational pathology, it is vital to explore the extraction of subvisual insights from the data via a *latent* representation which may not be apparently perceivable. For instance, CPath models can potentially help with identifying cellular communities (94,95) which can aid subvisual features extraction.

### 3.5 Interpretability and explainability

Many ML approaches, and notably DL methods, are known to be *black-box* approaches, making it hard or impossible to understand how the model reaches a decision. There are very few studies in the literature that discuss interpretability, which is typically via substitute measures like heat maps and feature maps. While the lack of interpretability and explainability is a common challenge for DL domain, it has a more pronounced impact in pathology since the decision making can influence diagnosis, prognosis, treatment planning and drug discovery roadmaps. More research is needed towards incorporation of interpretability and explainability of CPath models, be it through substitute measures or development of approaches that can help human interpretation.

## 3.6 Generalizability and artificial general intelligence (AGI) for pathology

The generalizability of an algorithm in computational pathology is a key feature to process data from the *wild* (i.e. real-world) due to heterogeneity of scanners, sites, tissue types and the ML task(s) at hand. To advance the march towards generalizability, we also need CPath systems to have some level of artificial general intelligence (AGI). Though AGI is a distant and ambitious goal, as per current technological developments, we certainly need the development of comprehensive CPath systems that can analyze heterogenous data to handle a variety of pathology tasks (e.g. tumor segmentation & grading, nuclei analysis, survival analysis, stain analysis etc). Current AI solutions in CPath are typically designed for specific tasks and for specific tissue types, requiring retraining and often development from scratch when dealing with new tasks or new tissue types. CPath systems with AGI-like capability need to integrate such solutions in a unified system for a real-world impact on pathology practice.

## 3.7 The role of challenge contests

The computational pathology research has observed tremendous progress over the last decade, owing to the availability of larger sized datasets, compute resource, DL models, increased uptake of digital pathology and continuous efforts of the research community. One such effort is notably observed through a variety of challenge competitions (e.g. CAMELYON, PANDA, GLAS and BACH). To continue the advancement of CPath as a discipline, we need to consider organizing similar high-quality challenge competitions on larger problems which a focus on multi-modal data analysis, federated data analysis, generalizability, OOD detection, learning with abstinence, robustness analysis and AGI solutions. Such competitions will not only propel the development of new algorithms, but they are also likely to deliver solutions that have stronger robustness and generalizability.

## 3.8 Causality

Current ML approaches in CPath are associative by design in that they correlate input variables to target variables and do not explicitly model interventions or counterfactuals (96,97). For example, although existing CPath models can predict mutation status from an image, they do not inform whether morphological features associated with the prediction are indeed a result of the mutation or not. Typically, outcome data in CPath models is obtained from observational studies – either from pathology laboratories or from clinical trials – and using this data for modelling counterfactuals presents a number of interesting challenges and opportunities for future research. We conjecture that causal modeling with an ability to estimate individual treatment effects will lead to a mechanistic understanding of predictive image-based CPath biomarkers.

## 3.9 The real test of the algorithms

In order to test the limits of CPath algorithms and bridge the gap between their estimated and true performance, we propose a few ideas below for future work to make progress on the translational front.

### 3.9.1 Concordance studies

While many AI solutions for CPath typically compare performance on quantitative accuracy of the proposed solution against a pathologist, systematic concordance and discordance studies (similar to the IBM Watson for Oncology (98)) are lacking that compare clinical decision making against the algorithm's decision making and not just for individual sub-tasks (e.g. segmentation).

### 3.9.2 Computer-assisted pathology (CASP)

The objective of many SOTA CPath algorithms is to conduct automatic analysis. Though automatic analysis promises to relieve the clinical workflow owing to the tedious and laborious nature of manual analysis, it is a loftier and distant goal to accomplish since there are several challenges. On the other hand, we postulate that the development of computer-assisted pathology (CASP) systems is a lower-hanging fruit and needs to be part of the near-future focus. The CASP systems can focus on the development of interactive, well-integrated, and well-designed user interface systems for performance of common tasks (e.g., nuclear statistics estimation, TILs analysis, ki67 hotspot estimation etc). Some initial work on CASP development indicates increased efficiency and performance in clinical practice (99,100).

### 3.9.3 Reproducibility and repeatability

As discussed earlier in Section 2.3.5, there is a lack of reproducibility and repeatability studies in computational pathology. In theory, the computational approaches promise to be more reproducible since they do not suffer from tiredness and laborious nature of the task. However, there is a lack of evidence and such studies are needed to investigate the repeatability and reproducibility of AI solutions in CPath and to demonstrate their robustness.

### 3.9.4 Guidelines and use cases for AI in cellular pathology

The rapid progress of AI in cellular pathology during the last decade is in sharp contrast to the lack of clear guidelines and use cases (*datasets*) on how to make use of the AI solutions in cellular pathology. Without such guidelines and use cases and bias free benchmarked datasets of sufficient size, the AI solutions will likely continue to be confined to research use hence the need for the development of such guidelines and use cases. A key initiative in this direction is the formation of AI in Anatomic Pathology Workshop by the College of American Pathologists (CAP) to develop use cases for artificial intelligence in pathology (101).

### 3.9.5 Outcome based stratification

Histological subtypes for various different types of cancers are often based on a combination of morphological and architectural patterns signifying the different types and the degree of malignancy. CPath research has generally followed suit in developing AI models for matching the existing subtypes. However, the problem is confounded by the fact that some tumors have a large number of subtypes: for instance, salivary gland tumors have 36 subtypes according to the WHO classification while the sarcomas have more than 100 subtypes. In addition, existing histological subtypes face two major challenges: (a) the larger the number of histological subtypes, the more difficult it may be for pathologists to agree on those and hence the more likely the higher inter-/intra- observer variation due to the inherent subjective nature of the subtype assignment via visual examination and the more difficult it is to obtain datasets with high quality ground-truth for training CPath models; and (b) the subtypes may not necessarily be based on outcomes making their usefulness questionable. The disciplines of pathology and oncology, and consequently the cancer patients, stand to benefit from steering the focus of CPath based histological subtyping towards outcome-based subtyping.

### 3.9.6 Turing test for computational pathology

The emergence of new CPath developments happening at pace can lead to the creation of a hype atmosphere, but it is imperative to balance the hype from the hope to enhance the acceptance and trust of these new developments. We propose to design a Turing test for pathology, similar to the one proposed here for cancer (102), whose objective will be to observe how AI solutions can assist in decision making for diagnosis, prognosis and treatment planning. We realize that the design of such a test will be a long process, but initially the test can be designed for an individual task (e.g. cancer detection, cancer grading, TILs grading etc) and later on the test can be evolved for the ability to handle a group of tasks along the lines of AGI as discussed above.

### 3.9.7 Wider collaboration and consultation

We believe that a closer collaboration between clinical, academic, industrial and patient/public stakeholders is the need of time. Recent initiatives in Europe such as the PathLAKE, NPIC and BIGPICTURE projects, with these projects involving all four types of stakeholders, are expected to lead the way in delivering guidelines and standards for the development and deployment of CPath models in practice. In particular, standards and guidelines are needed for: (a) storing, archiving, reading, collection, curation and sharing of WSIs with linked image-level and patient-level (e.g., clinical and genomic) annotations, (b) robust validation and generalizability of CPath models and (c) deployment, readouts and interpretability of AI algorithms.

## 3.10 Downskilling concerns

There is concern among some quarters that a new breed of pathologists using AI algorithms may gradually become so reliant on algorithms that they may lose their ability to recognize some nuanced histological patterns that they may have picked otherwise. An example of Google Maps is given where a couple of phenomena have been observed: (a) the users seem to be increasingly relying on the app's ability to find the optimal route for drivers and even reroute the driver when they miss a turn and (b) most users seem to be not too keen to remember or try to understand the routes followed by them. Although it is a bit too early to remark on the likelihood of this eventuality in case of cellular pathology, this is a predictable consequence of the technology and one which will need to be addressed through training and quality assurance.

The future of CPath is bright as long as the CPath community can bridge the gap between the estimated performance of CPath models and their true performance *in the wild* in the march towards their successful launch into clinical and pharmaceutical orbits and staying in the orbits in the long run.

**Supplementary Materials**

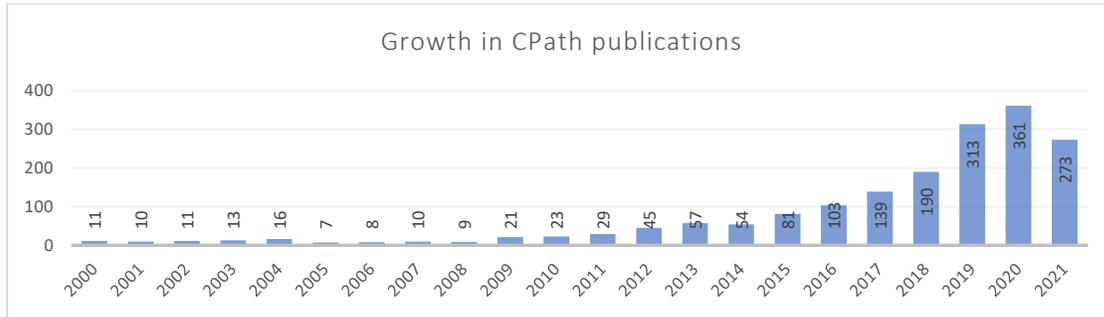

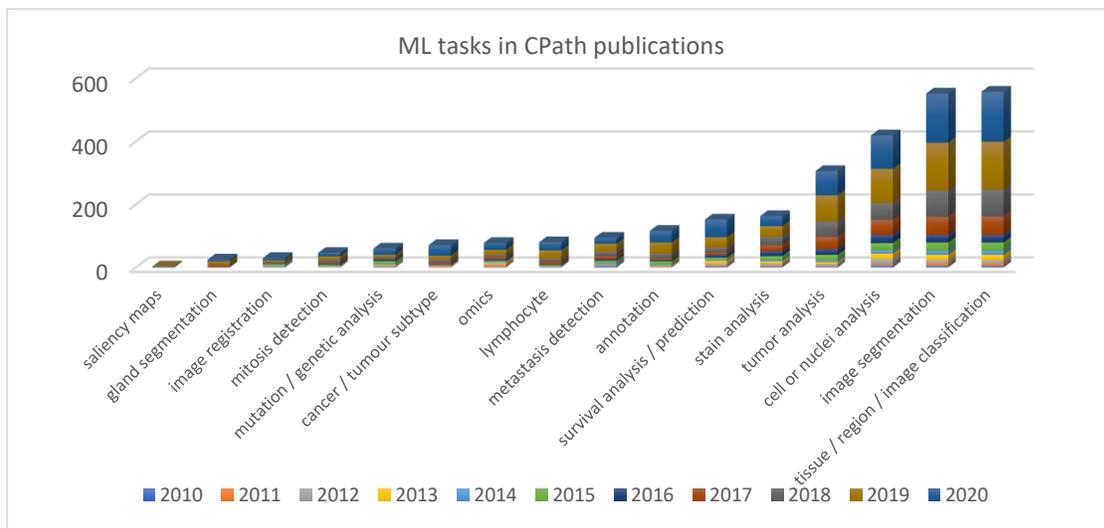

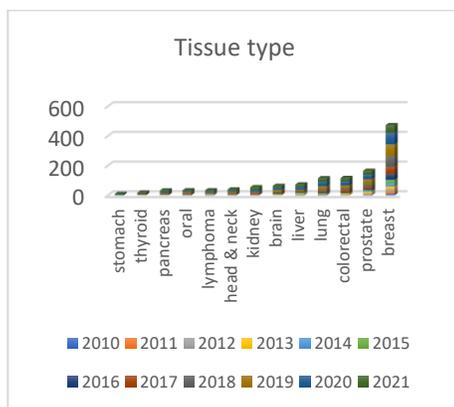

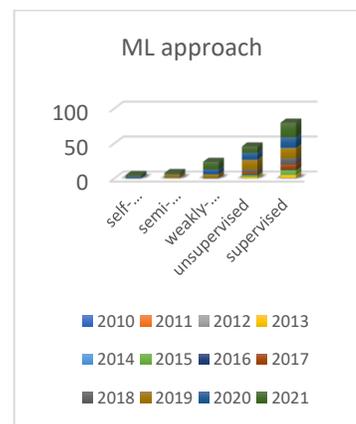

*Figure S1*: Computational pathology research trends analysis: (a) the number of papers published from year 2010 onwards; (b) the spread of various tasks studied; (c) the spread of various cancers studied; (d) the distribution of machine learning approach adopted by the community.